\documentclass[a4paper,fleqn,usenatbib]{mnras}
\usepackage{txfonts}

\usepackage[T1]{fontenc}
\usepackage{ae,aecompl}
\usepackage{graphicx}
\usepackage{color} %\definecolor{red}{rgb}{1.0,0.0,0.0}
\usepackage{pbox} %new line in table

\definecolor{gray}{RGB}{150,150,150}
\newcommand{\degree}{$^{\circ}~$}

\title[Summer fireworks on comet 67P]{Summer fireworks on comet 67P}
\author[J.-B. Vincent et al.]{
\parbox{\textwidth}{J.-B. Vincent$^{1}$\thanks{E-mail: \texttt{vincent@mps.mpg.de}}
M. F. A'Hearn$^{2}$
Z.-Y. Lin$^{3}$
M. R. El-Maarry$^{4}$
M. Pajola$^{5,6}$
H. Sierks$^{1}$
C. Barbieri$^{7}$
P. L. Lamy$^{8}$
R. Rodrigo$^{9, 10}$
D. Koschny$^{11}$
H. Rickman$^{12, 13}$
H. U. Keller$^{14}$
J. Agarwal$^{1}$
M. A. Barucci$^{15}$
J.-L. Bertaux$^{16}$
I. Bertini$^{6}$
S. Besse$^{17}$
D. Bodewits$^{2}$
G. Cremonese$^{18}$
V. Da Deppo$^{19}$
B. Davidsson$^{20}$
S. Debei$^{21}$
M. De Cecco$^{22}$
J. Deller$^{1}$
S. Fornasier$^{15}$ 
M. Fulle$^{23}$
A. Gicquel$^{1}$
O. Groussin$^{8}$
P. J. Guti\'errez$^{24}$
P. Guti\'errez-Marquez$^{1}$ 
C. G\"uttler$^{1}$ 
S. H\"ofner$^{1, 14}$
M. Hofmann$^{1}$
S. F. Hviid$^{25}$
W.-H. Ip$^{3, 30}$ 
L. Jorda$^{8}$ 
J. Knollenberg$^{25}$
G. Kovacs$^{1, 28}$
J.-R. Kramm$^{1}$
E. K\"uhrt$^{25}$
M. K\"uppers$^{17}$
L. M. Lara$^{24}$
M. Lazzarin$^{7}$
J. J. Lopez Moreno$^{24}$
F. Marzari$^{7}$
M. Massironi$^{27}$
S. Mottola$^{25}$ 
G. Naletto$^{28, 19, 6}$ 
N. Oklay$^{1}$
F. Preusker$^{25}$
F. Scholten$^{25}$ 
X. Shi$^{1}$
N. Thomas$^{4}$
I. Toth$^{29, 8}$
C. Tubiana$^{1}$}
\\~\\
\parbox{\textwidth}{
$^{1}$Max-Planck Institut f\"ur Sonnensystemforschung, Justus-von-Liebig-Weg, 3 37077 Goettingen, Germany;
$^{2}$Department for Astronomy, University of Maryland, College Park, MD 20742-2421, USA; 
$^{3}$Institute of Astronomy, National Central University, 32054 Chung-Li, Taiwan; 
$^{4}$Physikalisches Institut, Sidlerstrasse 5, University of Bern, CH-3012 Bern, Switzerland;
$^{5}$NASA Ames Research Center, Moffett Field, CA 94035, USA;
$^{6}$Centro di Ateneo di Studi ed Attivit\'a Spaziali "Giuseppe Colombo" (CISAS), University of Padova, Via Venezia 15, 35131 Padova, Italy;
$^{7}$Department of Physics and Astronomy "G. Galilei", University of Padova, Vic. Osservatorio 3, 35122 Padova, Italy;
$^{8}$Aix Marseille Universit\'e, CNRS, LAM (Laboratoire d'Astrophysique de Marseille) UMR 7326, 13388, Marseille, France;
$^{9}$Centro de Astrobiologia (INTA-CSIC), European Space Agency (ESA), European Space Astronomy Centre (ESAC), P.O. Box 78, E-28691 Villanueva de la Canada, Madrid, Spain;
$^{10}$International Space Science Institute, Hallerstrasse 6, 3012 Bern, Switzerland;
$^{11}$Scientific Support Office, European Space Agency, 2201 Noordwijk, The Netherlands;
$^{12}$PAS Space Research Center, Bartycka 18A, 00716 Warszawa, Poland; 
$^{13}$Department of Physics and Astronomy, Uppsala University, Box 516, 75120 Uppsala, Sweden;
$^{14}$Institute for Geophysics and Extraterrestrial Physics, TU Braunschweig, 38106 Braunschweig, Germany; 
$^{15}$LESIA, Observatoire de Paris, CNRS, UPMC Univ Paris 06, Univ. Paris-Diderot, 5 Place J. Janssen, 92195 Meudon Pricipal Cedex, France;
$^{16}$LATMOS, CNRS/UVSQ/IPSL, 11 Boulevard d'Alembert, 78280 Guyancourt, France;
$^{17}$ESA/ESAC, Camino Bajo del Castillo s/n, Ur. Villafranca del Castillo 28692 Villanueva de la Canada, Madrid, Spain;
$^{18}$INAF Osservatorio Astronomico di Padova, Vicolo dell'Osservatorio 5, 35122 Padova;
$^{19}$CNR-IFN UOS Padova LUXOR, Via Trasea 7, 35131 Padova, Italy;
$^{20}$NASA Jet Propulsion Laboratory, 4800 Oak Grove Drive, Pasadena, CA 91109, USA;
$^{21}$Department of Industrial Engineering University of Padova Via Venezia, 1, 35131 Padova;
$^{22}$University of Trento, via Sommarive, 9, 38123 Trento, Italy;
$^{23}$INAF - Osservatorio Astronomico di Trieste, via Tiepolo 11, 34143 Trieste, Italy;
$^{24}$Instituto de Astrofisica de Andalucia-CSIC, Glorieta de la Astronomia, 18008 Granada, Spain; 
$^{25}$Institute of Planetary Research, DLR, Rutherfordstrasse 2, 12489 Berlin, Germany; 
$^{26}$Budapest University of Technology and Economics, Department of Mechatronics, Optics and Engineering Informatics, Muegyetem rkp 3, Budapest, Hungary;
$^{27}$Dipartimiento di Geoscienze, University of Padova, via Granedigo 6, 35131, Padova, Italy;
$^{28}$Department of Information Engineering, University of Padova, Via Gradenigo 6/B, 35131 Padova, Italy;
$^{29}$Observatory of the Hungarian Academy of Sciences, PO Box 67, 1525 Budapest, Hungary;
$^{30}$Space Science Institute, Macau University of Science and Technology, Macau}
}

\date{Accepted 2016 September 20. Received 2016 August 26; in original form 2016 June 17}

\pubyear{2016}

\begin{document}
\label{firstpage}
\pagerange{\pageref{firstpage}--\pageref{lastpage}}
\maketitle
%-------------%

\begin{abstract}
During its two years mission around comet 67P/Churyumov-Gerasimenko, ESA's Rosetta spacecraft had the unique opportunity to follow closely a comet in the most active part of its orbit. Many studies have presented the typical features associated to the activity of the nucleus, such as localized dust and gas jets. Here we report on series of more energetic transient events observed during the three months surrounding the comet's perihelion passage in August 2015.

We detected and characterized 34 outbursts with the Rosetta cameras, one every 2.4 nucleus rotation. We identified 3 main dust plume morphologies associated to these events: a narrow jet, a broad fan, and more complex plumes featuring both previous types together. These plumes are comparable in scale and temporal variation to what has been observed on other comets. 

We present a map of the outbursts source locations, and discuss the associated topography. We find that the spatial distribution sources on the nucleus correlates well with morphological region boundaries, especially in areas marked by steep scarps or cliffs. 

Outbursts occur either in the early morning or shortly after the local noon, indicating two potential processes: Morning outbursts may be triggered by thermal stresses linked to the rapid change of temperature; afternoon events are most likely related to the diurnal or seasonal heat wave reaching volatiles buried under the first surface layer. In addition, we propose that some events can be the result of a completely different mechanism, in which most of the dust is released upon the collapse of a cliff.
\end{abstract}

   % MNRAS
  \newpage~
%
%________________________________________________________________
%---------------------------%
% INTRODUCTION
%---------------------------%
\section{Introduction}

The OSIRIS cameras on board ESA's Rosetta spacecraft have monitored the activity of comet 67P-Churyumov-Gerasimenko (67P) across varying heliocentric distances (4 AU to 1.24 AU) and different seasons on the nucleus (sub solar latitude between +45 and -55 degrees). Previous publications focused particularly on coma features usually referred to as jets: collimated streams of dust and gas arising from the nucleus. The foot prints of these features on 67P, their migration with the seasons and heliocentric distance, their relation to topography, their photometry, and possible formation mechanisms are described in details in \cite{lara2015, lin2015, lin2016} and \cite{vincent2016}. 

\begin{figure}
   \centering
   \includegraphics[width=0.49\hsize]{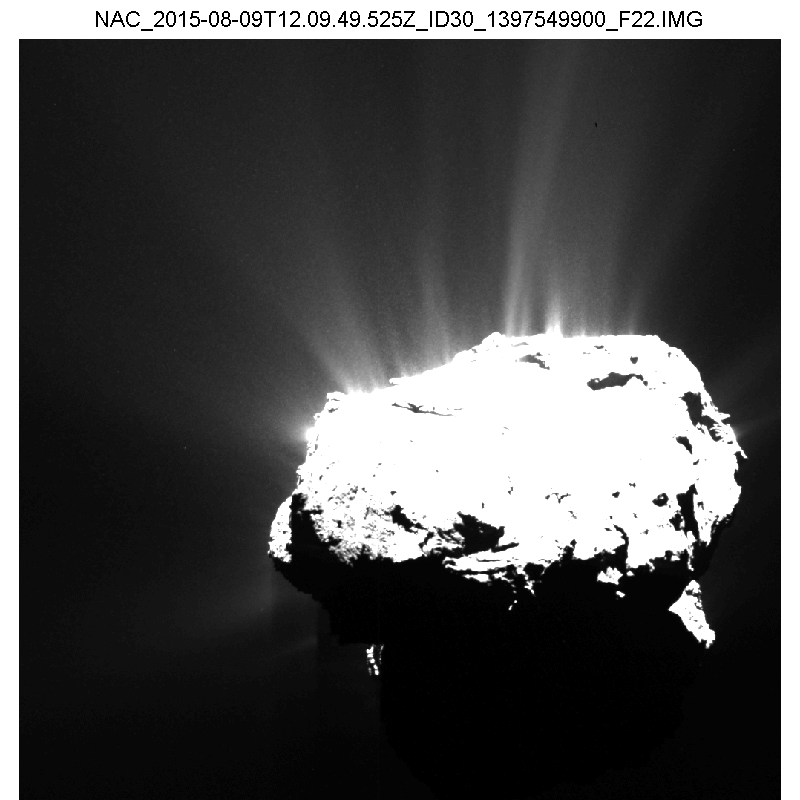}
   \includegraphics[width=0.49\hsize]{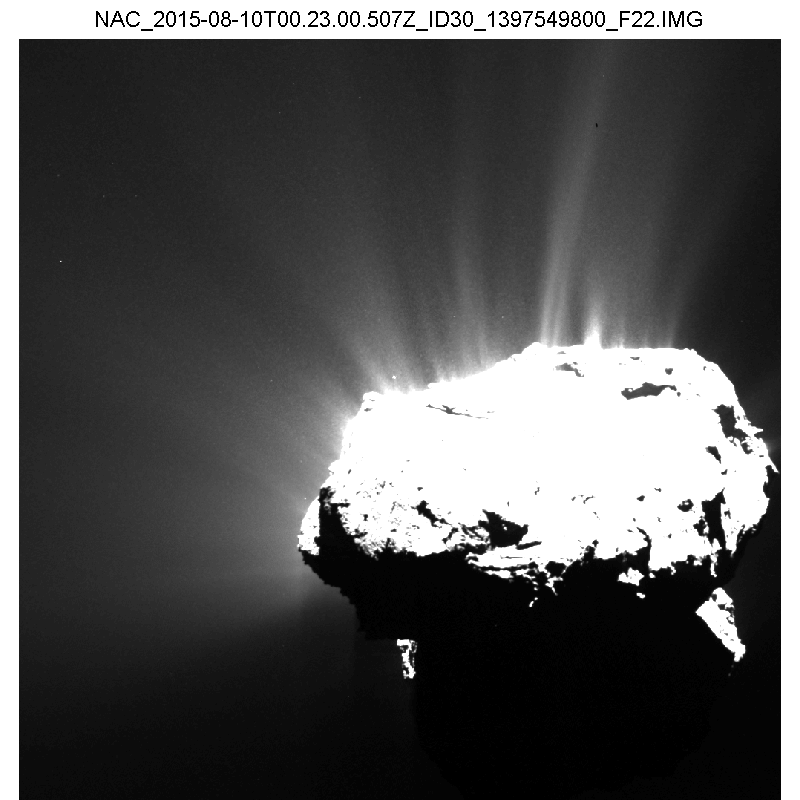}
   \caption{Example of two images acquired on 2015-08-09T12.09.49 (left) and 2015-08-10T00.23.00 (right), almost one rotation apart (Rotation period - images separation = 5min33s). Both images contrast is stretched to the same level (5\% of the same maximum brightness value). Field of view 1x1 degree, distance = 305 km, resolution = 5.7 m/px.}
    \label{fig:repeated_jets}
\end{figure}

One of the striking discoveries of Rosetta has been the clockwork repeatability of jets from one rotation to the next. Jets are very dynamic by nature, depending on the complex hydrodynamics of the gas and dust streams interacting with the local topography, and controlled by local thermal conditions. They grow and fade with the solar illumination as the nucleus rotates, but the same exact features can be observed from one rotation to the next. Figure \ref{fig:repeated_jets} shows an example of this phenomenon. This, of course, put constraints on the thermophysics and volatile content of active areas, which need to ensure the sustainability and repeatability of the jets we observed.

Long-lasting repetitive features are however not the only manifestation of activity on comet 67P. In this paper, we report on another types of events, much more transient in nature, which were observed most frequently around the summer months of 67P's southern hemisphere, i.e. from July to September 2015, when the comet reached its perihelion (13 August 2015, 1.24 AU).

\begin{figure}
   \centering
   \includegraphics[width=0.32\hsize]{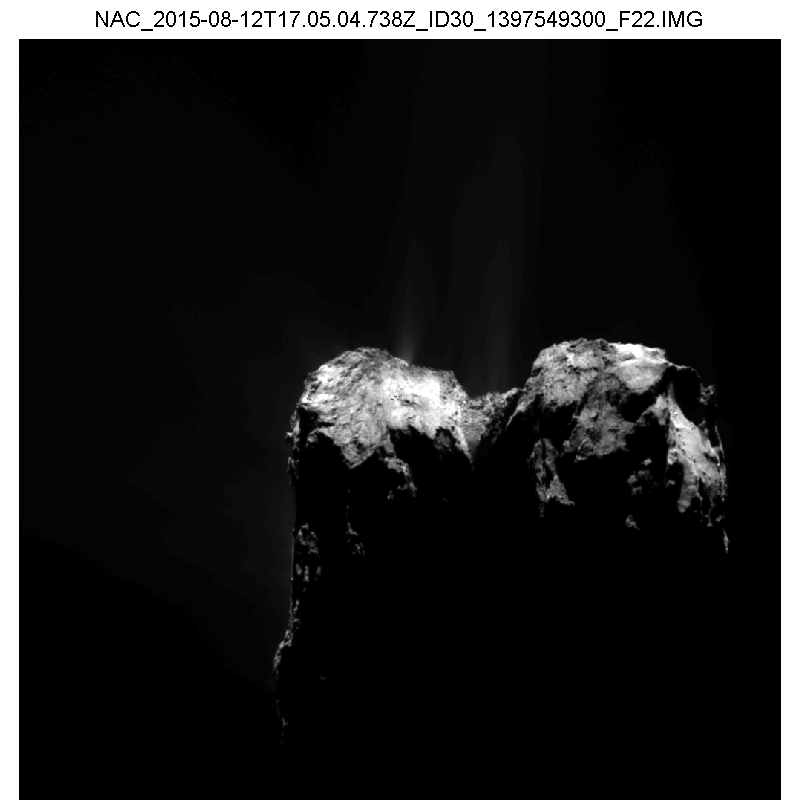}
   \includegraphics[width=0.32\hsize]{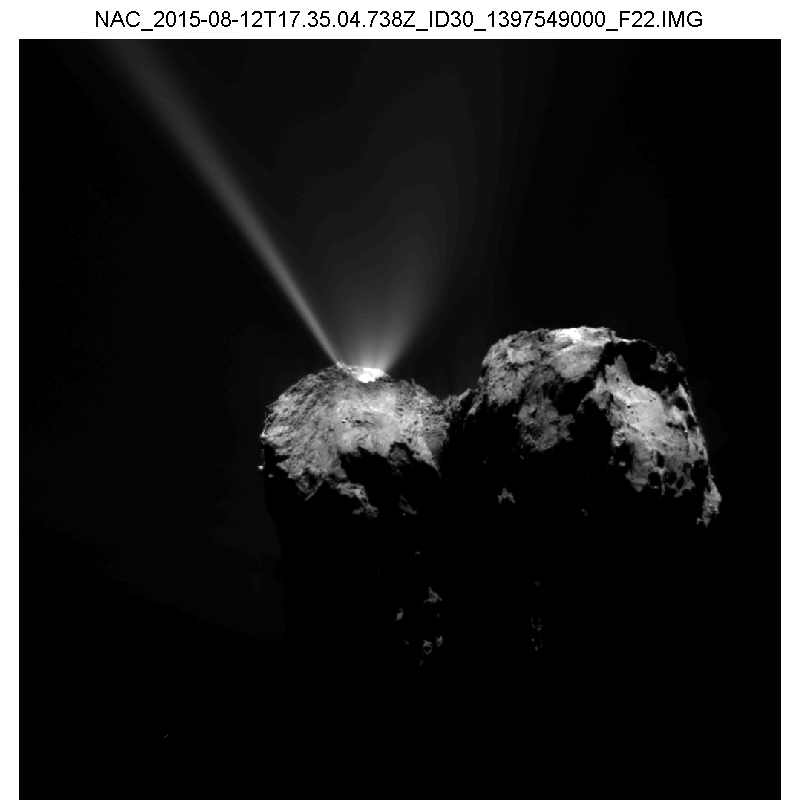}
   \includegraphics[width=0.32\hsize]{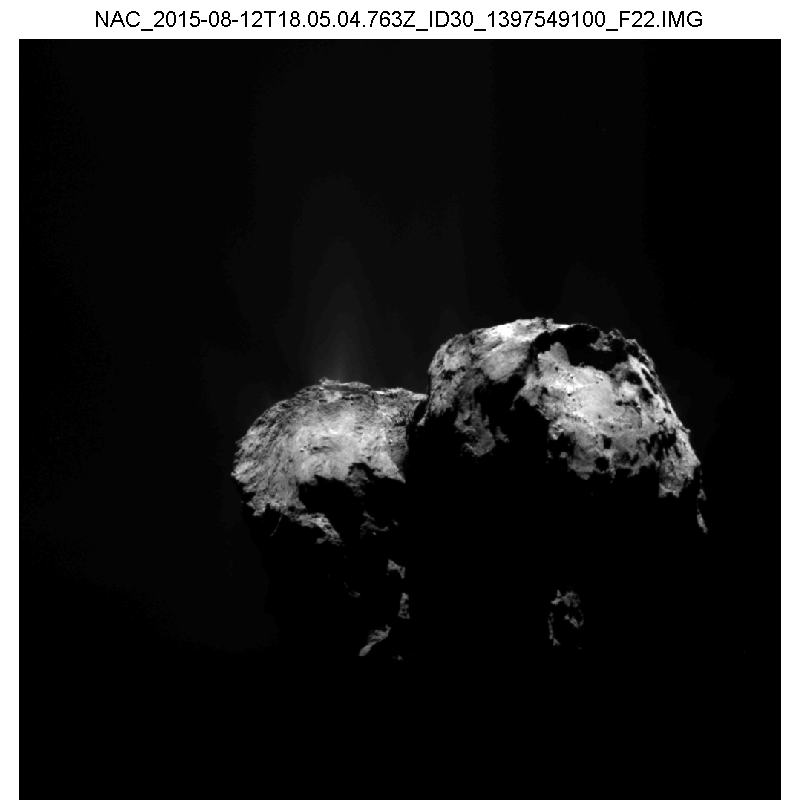}
   \caption{A transient event detected on the day of perihelion (12 August 2015). Images are separated by only 1/2 h, contrast not enhanced. Observations before and after the event show only faint activity from the outbursting area, while the image at 17:35 reveals a plume of material as bright as the nucleus, expanding at least 10km from the source. Field of view 1x1 degree, distance = 332 km, resolution = 6.1 m/px. Outburst \#14 in Table \ref{tab:outbursts}.}
    \label{fig:perihel_out}
\end{figure}

These events are characterized by the sudden and short release of a dust, sometimes collimated but not necessarily. While the typical jets are relatively faint (about 10\% of the nucleus surface brightness), the plumes ejected by these outbursts are usually as bright as the nucleus, and they can be detected in our images without enhancing the contrast. Contrary to the jets that last for several hours, most transient events are observed only once, indicating a lifetime shorter than the cadence of our images (between 5 and 30 min, depending on the observing sequence). One sequence showing a transient event is presented in Fig. \ref{fig:perihel_out}.

We report here our detection of these transient events, during a 3-months period surrounding the perihelion passage. In the following text transient events will alternatively be referred to as \textit{outbursts} to indicate their sudden and bright behavior, bearing in mind that they are many orders of magnitude fainter than typical cometary outbursts detected routinely by ground based observers for other comets.

%---------------------------%
% DATA AND METHODS
%---------------------------%
\section{Data and methods}

\subsection{Detection}
We used monitoring data acquired by the OSIRIS Narrow Angle and Wide Angle Cameras (NAC \& WAC), as well as Rosetta's navigation camera (NavCam) to increase our temporal coverage. Around perihelion, OSIRIS monitoring campaigns were run on a weekly basis, with a set of images acquired every 1/2 h for slightly longer than the current nucleus rotation period (12h18m10s at perihelion). After noticing the first outbursts in July 2015, we increased the cadence of images in each observation, and reduced the time between monitoring campaigns to a few days. In addition to the OSIRIS data, we also looked for transient events in the navigation images, acquired about every 4 hours during the whole mission.

To distinguish between outbursts and other short lived features, we established the following definition: An outburst is identified by a sudden brightness increase in the coma, associated to a release of gas and dust over a duration very short with respect to the rotation period of the nucleus. Typically detected in one image only, depending on the observing cadence. The dust plume is typically one order of magnitude brighter than the usual jets. We did not impose plume morphology as a criterion.

\begin{figure*}
   \centering
   \includegraphics[width=\hsize]{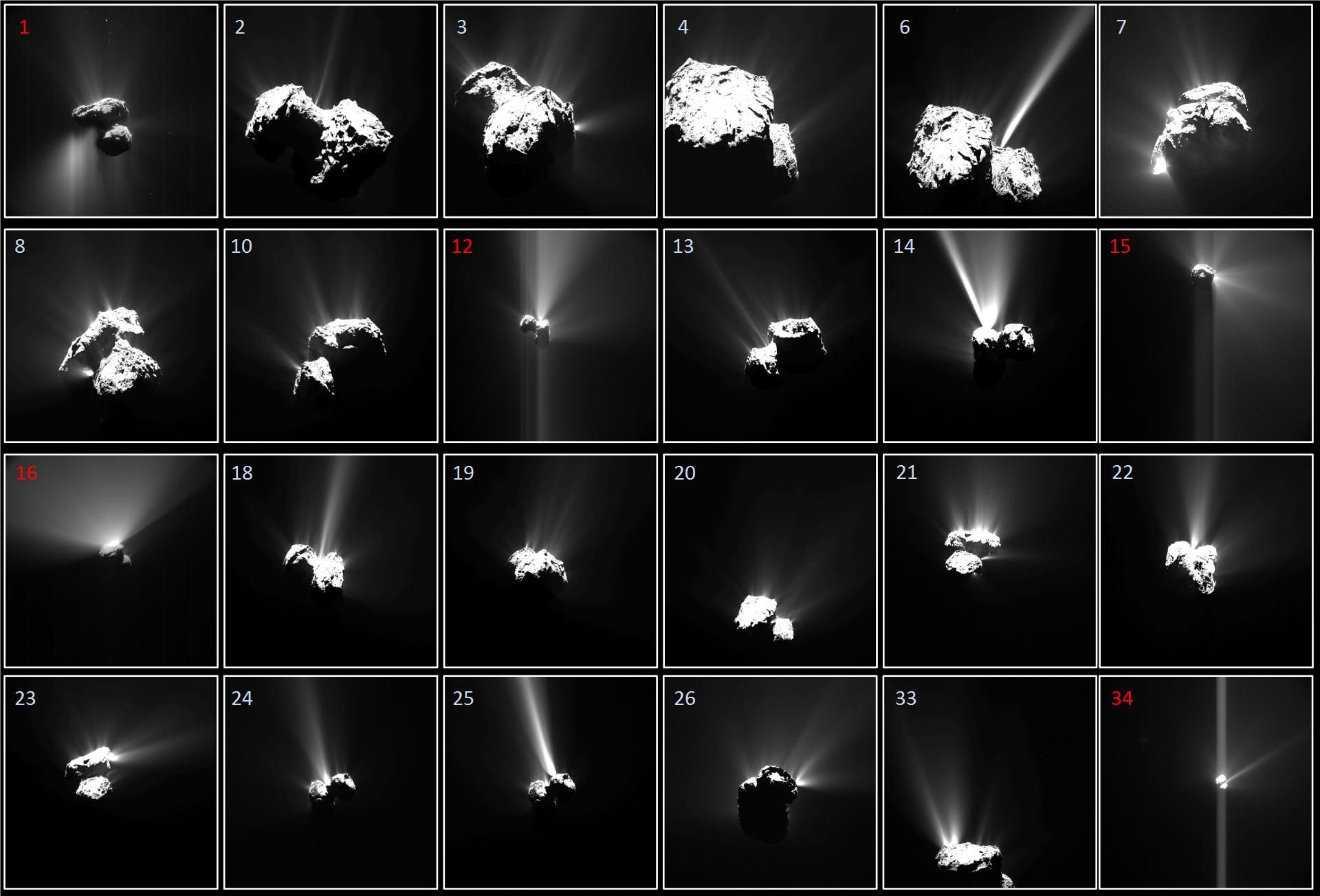}
   \caption{Mosaic of the brightest OSIRIS NAC (white) and NavCam (red) outbursts detected by Rosetta from July to September 2015. Observation details are give in \ref{tab:outbursts}. See Acknowledgments for detailed credit lines of the images.}
    \label{fig:mosaic}
\end{figure*}

Following this definition we identified 34 events in our data set, listed in Table \ref{tab:outbursts}. Among them, 26 were detected with OSIRIS NAC, 3 by OSIRIS WAC, and 5 by the NavCam. A visual catalog of the brightest evens is provided in Figure \ref{fig:mosaic}. A time line of these detections is given in Table \ref{tab:outbursts} and Figure \ref{fig:timeline}. 

\subsection{Source location}\label{sec:source_loc_inv}
We estimated the source location of each event with the three following techniques:

In most cases (for instance Fig. \ref{fig:perihel_out}), the ejected dust plume arises from one area in the field of view of our cameras. We measure directly the 2D coordinates of the source pixel in the image, and project this position in 3 dimensions on the most accurate shape model of 67P, obtained from stereo-photogrammetry for the Northern hemisphere \citep{preusker2015}, and stereo-photoclinometry for the southern hemisphere \citep{jorda2016}. The projection from image to shape is done using the spacecraft and comet reconstructed attitudes and trajectories provided by the SPICE library \citep{acton1996}. We verify this position by producing synthetic images from the orbital parameters and camera descriptions, and matching the the synthetic view with the original image. The largest uncertainty on this type of source inversion comes from the resolution of the images (3-6~m/px for the NAC, 15-32~40m/px for the WAC, and 15-83~m/px for the NavCam). This is the typical error for most of our observations.

If the source is not visible (typically just behind the horizon) but the plume is detected more than once (i.e. when acquiring multi-spectral images), we use the slightly different viewing geometry between the images to triangulate the source position. This technique is commonly used to find jet sources and described in details in \cite{vincent2016}. This leads to a maximum error smaller than 300~m on the surface, along the line of sight direction.

Finally, if the source is not visible and the plume is seen only in one image, then we can only roughly infer the source location. We noticed however that dust plumes released during outbursts events are often brighter than the nucleus very close to the source. We used this to constrain how far the source must lie beyond the horizon, and found a maximum uncertainty of about 10 degrees in latitude and longitude, equivalent to about 300~m on the surface.

Sources with a large uncertainty (9 out of 34) are indicated with an error ellipse on the map given in Fig. \ref{fig:maps}. All other positions have an uncertainty comparable to the size of the symbol used to mark the source location on the figure.

\subsection{Relative intensity of outbursts}\label{sec:intensity}
Characterizing the ejected mass per outburst is challenging because of their transient nature. As we see the event in one image only during a given sequence, we can only know that all the material was ejected in the time frame covered by three images, like in Fig. \ref{fig:perihel_out}. We do not know if the image showing the dust plume was acquired shortly after the outburst started, or later in the process. Additionally, the plume often extends beyond our field of view and we do not know enough about the acceleration regime to be able to extrapolate the visible brightness profile.
However, when we had the chance to follow an outburst for more than one image (i.e. outburst \#26), we saw that the total duration of the event was at most a couple of minutes. This means any image showing an outburst is likely to have been acquired very shortly after the event started, and therefore most of the ejected material is still in the field of view. For the same reason, outbursts are likely to have been observed in a similar stage of development and by integrating their brightness over a large area we can ignore local variations due to the non-steady state of the plume.

For each event, we integrated the total radiance ($W.m^{-2}.sr^{-1}.nm^{-1}$) measured in a trapezoid box extending along the edges of the plume from a distance of 50 m to 500 m. The closest boundary was chosen to avoid capturing any remnant signal from the illuminated nucleus, in case the ejected dust cloud is not optically thick. The furthest boundary is arbitrarily set 10 times further to ensure that we capture enough material to draw a meaningful comparison, independently of the plume morphology. This means we typically integrated about 5000 NAC pixels per outburst. The integrated radiance is converted to luminosity ($W.nm^{-1}$) by a multiplication with the factor $4\pi cdist^2$, with $cdist$ the distance between Rosetta and the comet. This luminosity is finally normalized to the brightest outburst we observed in this period (\#14: 12 August 2015, a.k.a. "Perihelion Outburst"). The luminosity of this event, integrated in the window described above, was measured to be $1.18 \times 10^{13}~W$ at 649.2~nm (OSIRIS NAC Orange filter 'F22').

The perihelion outburst was the strongest ever observed, at least an order of magnitude above most of the other events. A more average outburst was observed on the 29$^{th}$ of July (\#06 in Table \ref{tab:outbursts}). By chance, most Rosetta instruments were acquiring data at that time and the first results of this common analysis have been presented online shortly after the event (http://blogs.esa.int/rosetta/2015/08/11/comets-firework-display-ahead-of-perihelion). We modeled the photometric profile of the July 29 event using the approach described in \cite{knollenberg2015}. In short, we convert the radiance over an image area as described above, convert it to a dust cross section, and then to mass assuming a dust size distribution with a power law of -2.6. For this specific event we estimated an ejected flux of 60 to 260 $kg.s^{-1}$ for particles in the range 1-10 $\mu m$ or 1-50 $\mu m$. This is equivalent to 4-17\% of the total dust flux being ejected by the comet with its nominal activity at perihelion (1500 kg/s, \citet{fulle2016}). We know from other events and the cadence of our images that outbursts last less than 5min. Therefore the maximum mass of dust ejected by this event is in the order of 20 to 80 tons. 

This mass loss per outburst is comparable to what has been observed on at least one other comet. We have estimated the mass ejected in the outburst by 9P/Tempel 1 on 2005 Jul 2 \citep{ahearn2005} using the archived photometry from Deep Impact (Bastien et al, 2008, Deep Impact MRI Photometry of Comet 9P/Tempel 1 V1.0, DIF-C-MRI-5-TEMPEL1-PHOTOMETRY-V1.0, NASA Planetary Data System) and common assumptions about the relationship between brightness and dust. This outburst was ~2/3 the brightness of the ambient coma at all radii and the total mass was 300 tons ejected over 10min, therefore an outburst flux of 500 $kg.s^{-1}$, of the same order as the values we have found for 67P.

At the time of writing this paper the NavCam data is not calibrated. We indicate these events in our timeline but arbitrarily set their relative brightness to zero. Some did however saturate the NavCam CCD when the nucleus did not (i.e. outburst \#15, on 21/08/2016), so one can expect that after calibration they will appear at least as strong as the outbursts typically detected by OSIRIS.

\begin{table*}
\caption{Detected outbursts and locations of their sources}             
\label{tab:outbursts}      
\centering          
\begin{tabular}{c c c c c c c c c c c c}  
\hline\hline
Id & date & camera & hdist & cdist & lat      & lon       & Sun lat & Sun lon & Relative  & Type & Time since\\
 ~ &[UTC] &   ~    & [AU]  &  [km] &[\degree] & [\degree] &[\degree]&[\degree]& Luminosity&   ~  &Sunrise [h]\\
\hline                    
01 & 2015-07-10T02:10:18 & NavCam & 1.311 & 155.16 &  74 & 200 & -30.51 & 131.65 & 0.00\% & B & 3.62\\
02 & 2015-07-19T03:38:09 &    NAC & 1.281 & 180.00 & -24 & 296 & -35.70 & 292.44 & 2.43\% & A & 3.09\\
03 & 2015-07-26T20:22:42 &   NAC  & 1.261 & 168.00 & -36 &	75 & -39.94 & 316.99 & 7.07\% & B & 11.16\\
04 & 2015-07-27T00:14:29 &   NAC  & 1.261 & 168.00 & -31 & 333 & -40.03 & 204.51 & 1.94\% & A & 10.24\\
05 & 2015-07-28T05:23:43 &   WAC  & 1.259 & 180.87 &  -4 & 264 & -40.67 &  75.48 & 0.51\% & B & 10.73\\
06 & 2015-07-29T13:25:28 &   NAC  & 1.256 & 186.00 & -37 & 300 & -41.37 & 222.65 & 15.58\%& A & 3.69\\
07 & 2015-08-01T10:53:15 &   NAC  & 1.252 & 214.05 & -12 & 196 & -42.84 & 358.46 & 1.41\% & B & 10.51\\
08 & 2015-08-01T15:44:50 &   NAC  & 1.251 & 211.00 & -28 &	34 & -42.94 & 216.81 & 11.71\%& B & 10.68\\
09 & 2015-08-03T17:27:03 &   WAC  & 1.249 & 218.49 & -75 & 303 & -43.94 & 207.71 & 0.55\% & B & 9.97\\
10 & 2015-08-05T07:25:05 &   NAC  & 1.247 & 253.00 & -25 & 320 & -44.69 & 180.67 & 1.90\% & A & 10.32\\
11 & 2015-08-05T08:05:15 &   NAC  & 1.247 & 253.00 & -23 & 318 & -44.70 & 161.14 & 1.90\% & A & 10.47\\
12 & 2015-08-08T15:21:48 & NavCam & 1.244 & 303.00 & -30 &	51 & -46.17 &   7.96 & 0.00\% & C & 3.41\\
13 & 2015-08-09T09:15:14 &   NAC  & 1.244 & 304.00 & -30 & 298 & -46.48 & 205.71 & 3.05\% & A & 9.94\\
14 & 2015-08-12T17:21:20 &   NAC  & 1.243 & 332.00 & -30 &	58 & -47.81 &  26.46 & 100.00\% & C	& 3.32\\
15 & 2015-08-21T09:44:53 & NavCam & 1.247 & 330.00 & -32 & 227 & -50.54 &  52.98 & 0.00\% & B & 10.61\\
16 & 2015-08-22T06:47:04 & NavCam & 1.248 & 336.00 & -40 & 168 & -50.75 & 157.23 & 0.00\% & C & 3.15\\
17 & 2015-08-22T23:46:21 &   WAC  & 1.249 & 334.00 & -25 & 316 & -50.91 &  19.88 & 0.29\% & B & 11.60\\
18 & 2015-08-23T01:39:38 &   NAC  & 1.249 & 334.35 & -53 & 292 & -50.93 & 324.60 & 12.53\%& A & 3.33\\
19 & 2015-08-23T15:12:48 &   NAC  & 1.251 & 340.17 & -23 & 314 & -51.05 & 287.77 & 5.46\% & A & 3.28\\
20 & 2015-08-26T07:51:04 &   NAC  & 1.254 & 417.00 & -41 &	42 & -51.55 & 194.51 & 5.57\% & B & 10.43\\
21 & 2015-08-27T22:58:04 &   NAC  & 1.257 & 403.55 &  -8 & 321 & -51.80 & 128.16 & 2.71\% & B & 10.76\\
22 & 2015-08-28T02:29:21 &   NAC  & 1.257 & 403.8  & -21 &	24 & -51.82 &  24.94 & 29.04\% & C & 3.07\\
23 & 2015-08-28T10:10:57 &   NAC  & 1.258 & 410.29 & -31 & 229 & -51.86 & 159.42 & 69.84\% & B & 3.63\\
24 & 2015-09-05T08:50:02 &   NAC  & 1.276 & 436.07 & -15 &	26 & -52.31 & 325.22 & 27.89\% & C & 11.63\\
25 & 2015-09-05T09:00:02 &   NAC  & 1.276 & 435.4  & -31 & 330 & -52.31 & 320.33 & 66.73\% & C & 3.14\\
26 & 2015-09-10T08:59:49 &   NAC  & 1.291 & 317.9  & -25 &	67 & -52.04 &  33.74 & 1.83\% & C & 3.33\\
27 & 2015-09-10T13:06:14 &   NAC  & 1.292 & 317.41 & -23 & 292 & -52.03 & 272.97 & 6.47\% & A & 3.22\\
28 & 2015-09-10T13:36:14 &   NAC  & 1.292 & 317.38 & -21 & 307 & -52.02 & 258.27 & 9.44\% & C & 3.46\\
29 & 2015-09-10T14:11:15 &   NAC  & 1.292 & 317.34 & -15 &	10 & -52.02 & 241.10 & 7.97\% & B & 11.07\\
30 & 2015-09-10T18:57:41 &   NAC  & 1.292 & 317.54 & -15 &	10 & -52.00 & 100.71 & 4.37\% & A & 9.93\\
31 & 2015-09-10T19:27:41 &   NAC  & 1.292 & 317.6  & -30 & 286 & -52.00 &  86.00 & 7.45\% & C & 10.82\\
32 & 2015-09-12T09:41:00 &   NAC  & 1.298 & 329.89 & -12 & 318 & -51.82 &  41.69 & 15.96\%& C & 11.44\\
33 & 2015-09-14T18:47:00 &   NAC  & 1.306 & 316.29 & -25 & 198 & -51.49 & 161.29 & 35.21\%& C & 3.36\\
34 & 2015-09-26T12:03:32 & NavCam & 1.356 & 817.64 & -40 & 307 & -48.86 & 149.76 & 0.00\% & A & 10.47\\
\hline                  
\end{tabular}
\caption{Column description: Identification number of the outburst, date, camera which detected the event, 67P heliocentric distance, Rosetta cometocentric distance, latitude and longitude of the source, latitude and longitude of the sub-solar point, relative intensity with respect to the brightest event, type of outburst. Time since Sunrise is an estimation of the local time on the surface (see Sec. \ref{sec:local_time}: based on a 12.25~h rotation period, a time-since-sunrise = 3h indicate the local mid-day, while a time-since-sunrise = 12h is the end of the night/early morning.}
\end{table*}

%---------------------------%
% RESULTS
%---------------------------%
\section{Results}
\subsection{Morphological classification of the dust plumes}\label{sec:ob_morpho}
We distinguish at least three types of dust plumes morphology associated to outbursts, from which we derived the following classification:
\begin{itemize}
\item Type A: They produce a very collimated jet which expands beyond our field of view (typically 10~km). They extend further away from the nucleus than the other types.
\item Type B: Broad plume, or wide dust fan. They expand much more laterally than radially when compared to type A plumes. 
\item Type C: Complex events, often combining both a narrow and a broad feature. To the best of our knowledge both features arise from the same source, within the error ellipse of our detection.
\end{itemize}

\begin{figure}
   \centering
   \includegraphics[width=0.3\hsize]{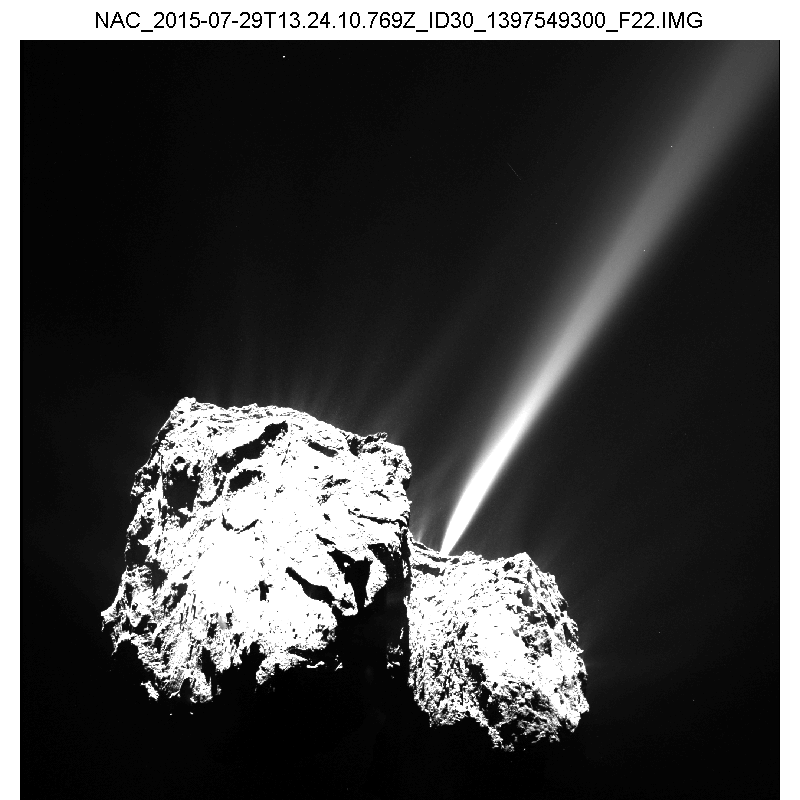}
   \includegraphics[width=0.3\hsize]{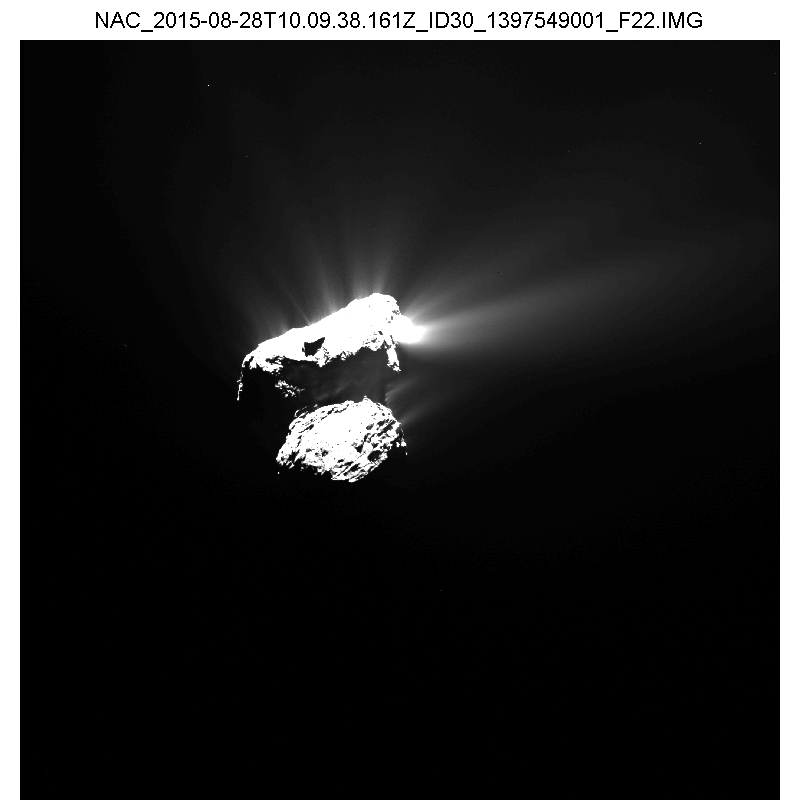}
   \includegraphics[width=0.3\hsize]{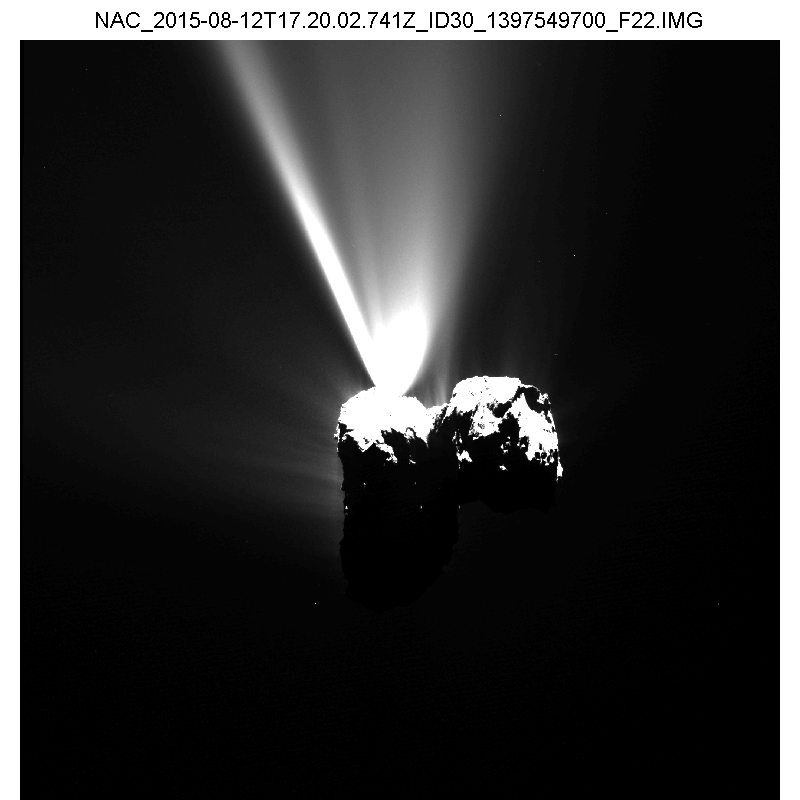}
   \caption{The three major morphologies of outbursts defined as Types A (jet), B (broad plume), and C (complex) in section \ref{sec:ob_morpho}. Outbursts \#06, \#23, and \#14 in Table \ref{tab:outbursts}.}
    \label{fig:ob_type}
\end{figure}

All morphologies seem equally probable. The type of each event is indicated in Table \ref{tab:outbursts} and an example of each is given in Figure \ref{fig:ob_type}. It is important to stress that this classification is purely morphological. It is not clear whether the 3 types correspond to different mechanisms or if they are different stages of a same process. 

We can constrain the dust velocity in these plumes by using the cadence of our images. For instance the plume associated to event \#25 extends by at least 8150 m (edge of NAC frame) and was not detected in the previous image acquired 10 min earlier. This implies that dust was ejected with a minimum velocity of 13 $m.s^{-1}$. This is at least one order of magnitude larger than the typical velocity of dust grains in 67P's jets (1 $m.s^{-1}$, \cite{lin2016}), indicative of more energetic events.

We tried to identify whether the different plume morphology reflects an evolutionary process in the outburst mechanism. We looked for morphological variations when the imaging cadence showed the same event in several images but could not observe any change of morphological type, only the expansion of the dust plume. Nonetheless, we cannot exclude an observational bias. 

If there is indeed an evolution, it seems that the most reasonable sequence would be type A -> type C -> type B:
\begin{itemize}
\item The event starts with some dust and gas being ejected at high velocity in a narrow plume (type A). This is indicative of a small source area, possibly confined.
\item As the outbursts unravels, the local surface is modified (collapse, or "eruption") and exposes a larger fraction of fresh material leading to the formation of a broader plume (type C).
\item Finally, the morphology of the source area has changed enough to not be able to collimate the initial narrow flow anymore, and only the broad plume survives (type B).
\end{itemize}

%---------------------------%
% --- Timeline
%---------------------------%
\subsection{Timeline of detected outbursts}
The summer outbursts presented here are not the first ones detected by the OSIRIS cameras on comet 67P. We did observe a large outburst in April 2014, at a distance of 4AU \citep{tubiana2015}. We did not detect any further event as Rosetta was closing in to the comet, down to a distance of 10km in October 2014. As the OSIRIS cameras were mapping the nucleus with high cadence imaging from July to October 2014, it is unlikely that we have missed an outburst in this period. 
The next event occurred in February 2015 at a distance of 2.5 AU and is described in \cite{knollenberg2015}. Although much smaller in scale that most of the other events, it is particularly interesting because it arose from an area that had been in the night for 5 hours when the outbursts occurred.

We did not detect any other event between February 2015 and the first summer event in July 2015. It is however possible that we missed some, as the high dust content in the vicinity of the spacecraft triggered safing events and a retreat to several hundred km from the nucleus, which prevented us from monitoring the activity as usual. Other instruments may have detected a few events in this time frame.

\begin{figure*}
   \centering
   \includegraphics[width=\hsize]{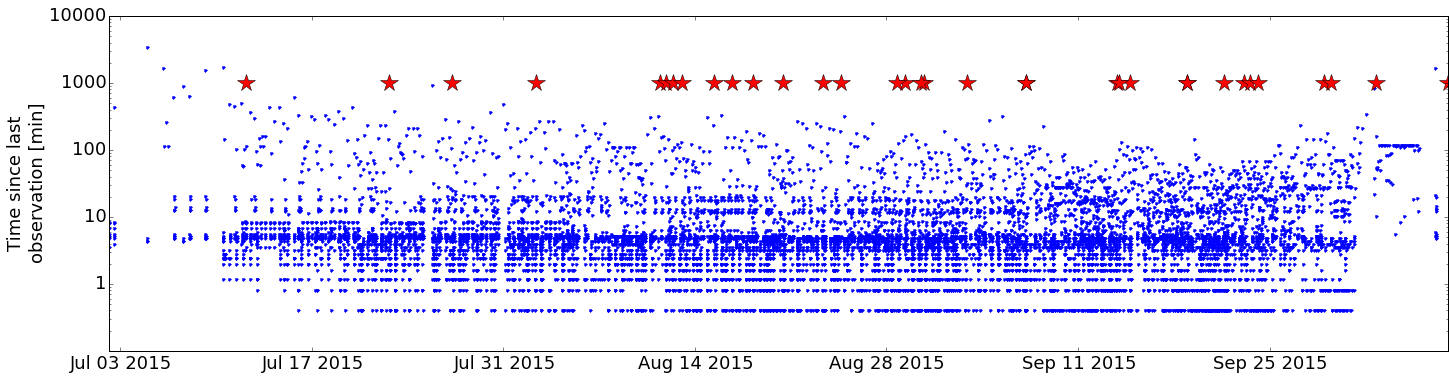}
   \caption{Time line of images and outbursts detection over the Summer 2015. Every blue dot represents an image acquisition, with the vertical axis showing how much time passed since the last observation. Red stars represent outburst detections.}
    \label{fig:timeline}
\end{figure*}

The OSIRIS cameras acquired 11807 images from the 1$^{st}$ of July to the 30$^{th}$ of September 2015, with an average time separation of 12 min. Among these observations We ran 12 dedicated outburst campaigns aimed at detecting and characterizing transient events with fast cadence imaging: 1 observation every 5min, for a few hours. Figure \ref{fig:timeline} shows the time line of our detections. We observed 34 events, about one every 1.27 days, i.e. every 2.37 comet rotation (period = 12.25h)

Comparing the time line of detected transient events with the cadence of our images gives us a hint of the completeness of our catalog. For instance, one can see that we may have missed several events in the first half of July 2015 due to poor time coverage. The same is true for the last week of September 2015 during which we did not observe for 24 consecutive hours. However the rest of the time line is densely covered with observations, and the gaps in outburst detection cannot be explained by lack of imaging. This is particularly true for the first half or August 2015 or the week around the 10$^{th}$ of September 2015 during which we did not detect any event in spite of continuous high-cadence monitoring. In addition to that, it happened several times that we observed the comet in consecutive rotations for 1 or 2 days during which only one event was detected. For instance, the full set of observations acquired around perihelion from 9$^{th}$ to 13$^{th}$ of August yielded only two events (9$^{th}$ and 12$^{th}$ of August).

Therefore it seems likely that the cadence of one outburst every 2.4 comet rotation is close to the real cadence of such events, and it may have implications on the related mechanism (see Section \ref{sec:mechanism}). This cadence is comparable to what has been reported for comet 9P/Tempel 1: 1 outburst every 3.3 days, i.e. every 2$^{nd}$ comet rotation (period = 40h).

Note that we consider here only events producing a dust feature detectable without enhancing the contrast of our images, i.e. comparable in brightness to the nucleus. We also see short lived jets in almost every sequence, and their number of detection increases with the imaging cadence. They are however very faint, typically less than 10\% of the nucleus brightness. Apart from their short duration, they behave comparably to all other nominal jet features.

%---------------------------%
% --- Source locations
%---------------------------%
\subsection{Source locations and local morphology}\label{sec:source_loc}
\subsubsection{Global map}
\begin{figure*}
   \centering
   \includegraphics[width=.9\hsize]{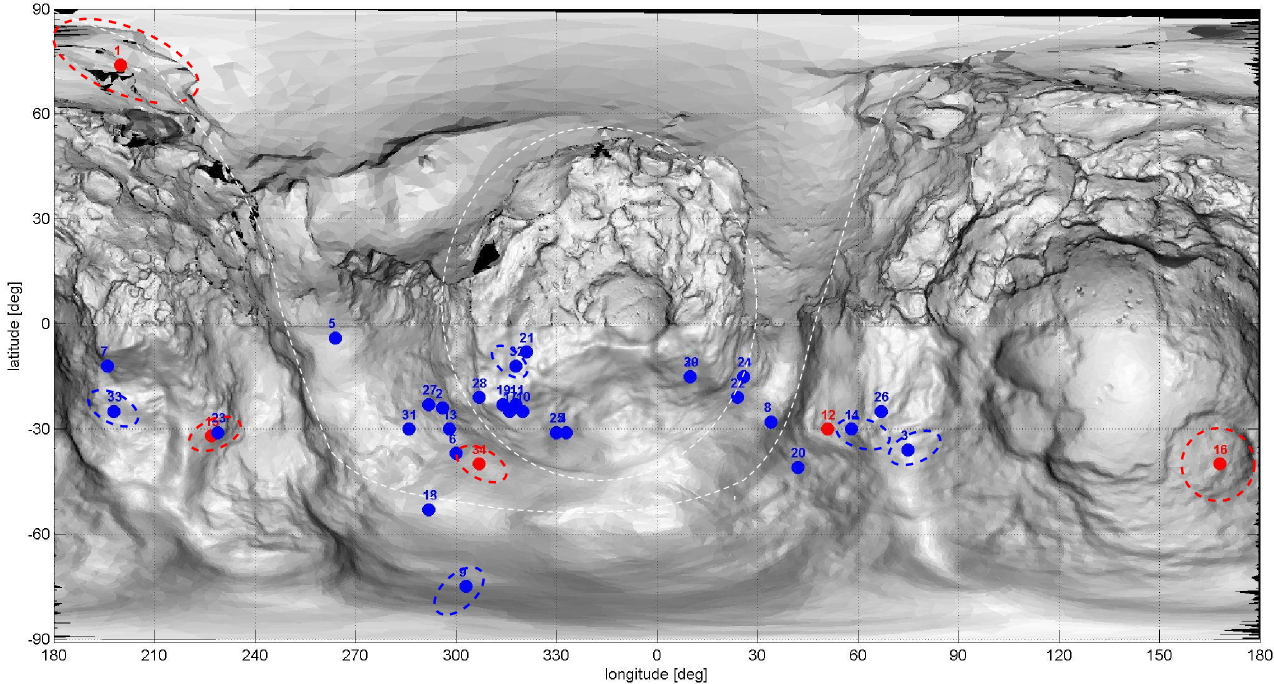}\\
   \includegraphics[width=.9\hsize]{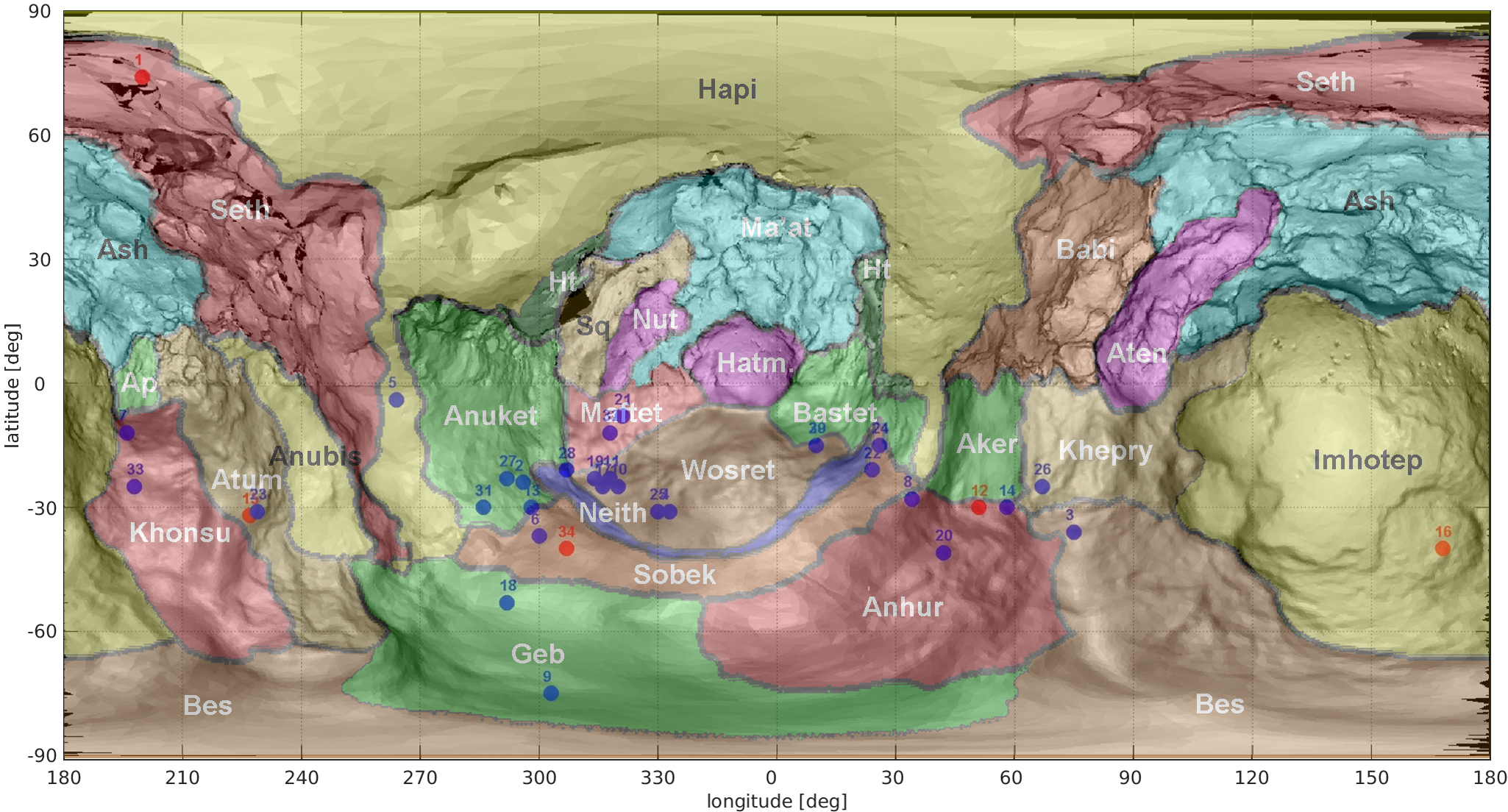}\\
   \caption{Maps of all summer outbursts detected by the OSIRIS cameras (blue dots) and Rosetta's NAVCAM (red dots). The top panel plots the sources over a topographic maps in which the gray shading represents the local gravitational slope (white=flat, black=vertical wall). Dotted ellipses represent the estimated uncertainty for the few outbursts whose source was not observed directly. Note that this map is a 2D representation of a bi-lobate, strongly concave object, and therefore presents significant distortions. To guide the reader, we indicated with white dashed lines the boundary of the two lobes: the map is centered on the small lobe, the big lobe covers the left-right-bottom edges of the map, and the contact area between the two lobes covers mainly the top of the map (regions Hapi, Neith, Sobek). The 3 main clusters of outbursts sources are located around longitudes 60 degrees (big lobe), 300 degree (southern neck), and 315 degree (small lobe).}
    \label{fig:maps}
\end{figure*}

Figure \ref{fig:maps} shows all outburst sources projected on a topographic map of 67P, and on a morphological map displaying the regions boundaries. 
All sources but one are located in the southern hemisphere, between 0 and -50 degrees of latitude, i.e. around the sub solar latitude for this period (it varied from -30 to -55 degrees). This is consistent with previous observations showing that active sources in general migrate with the Sun \citep{vincent2016, ip2016}.
Outburst sources are not evenly distributed along this latitude. We observe some clustering in three main areas: (1) The Anhur-Aker boundary (big lobe), (2) the Anuket-Sobek boundary (big lobe), and (3) the Wosret-Maftet boundary (small lobe). These areas are characterized by steep scarps, cliffs, and pits, which contrast with the overall flatter morphology of the Southern hemisphere. It is interesting to note that beyond those three areas, it seems like all outbursts sources are located close to morphological boundaries, i.e. areas where we observe discontinuities in the local terrain, either textural or topographic. Region boundaries are defined in \cite{elmaarry2015a,elmaarry2016}. This seems to indicate a link between morphology and outbursts, although it is not clear which one influences the other. We will discuss this further in Section \ref{sec:mechanism}.

\subsubsection{Local morphology}
The local morphology, especially at the time of the outburst, is more difficult to characterize for several reasons. All the events reported here were observed around perihelion while Rosetta was orbiting $\backsim$300 km from the nucleus. This distance corresponds to a spatial resolution of $\backsim$6 m/px, with the OSRIS NAC. Following perihelion passage, Rosetta undertook an excursion into the plasma tail at 1500 km away before approaching back slowly to the nucleus. As a result, high resolution images (< 1m/px) could only be acquired from February 2016 onward, when Rosetta went back to distances lower than 50 km. Additionally, the illumination conditions pre-perihelion were poor for the Southern hemisphere and prevented high resolution imaging. Therefore, for most of the outbursts only an "after" image, acquired 6 months after the event is available. When present, pre-perihelion images have too low spatial resolution for a confident quantitative comparison. Nevertheless, high resolution images of some of the source regions display interesting morphologies that worth mentioning. \par

Figure 7 shows a number of outbursts spots particularly in the Wosret and Anhur regions as well as notable regional boundaries such as those of Anuket/Sobek and Aker/Anhur. We observe that most sources fall on steep topographic structures such as scarps and pits, often displaying nearby talus deposits. The Wosret region (Fig. 7a) shows two distinctive morphological terrains one where many candidate outburst locations are observed in the pitted terrains as opposed to the heavily fractured and quasi-flat terrains. A link between outbursts and pitted terrains has been explored in details by \cite{belton2008} in their review of similar events of comet 9P, although the scale and distribution of pitted terrains is not the same as on 67P. Given the lower resolution of the images of 9P's surface (at best 10m/px) it is not cleat whether the outbursts carved out the pits, or if outbursts originate from a subsequent evolution, such has the collapse of the pit walls.

A number of outbursts appear to coincide remarkably with the boundary between Anuket and Sobek (the southern neck, Fig. 7b), and a number of terraces that also show considerable talus deposits. A closer look at the main cliffs of the southern part of the large lobe shows that many outburst locations appear to coincide either with some of the numerous niches and alcoves in the Anhur region (Fig. 7c). The Anhur region appears to be the most weakly consolidated region in the southern hemisphere as evident from it morphology, low slope in comparison to other cliff regions, and abundant boulders and debris \citep{elmaarry2016, pajola2016b}. Therefore, it is likely that this region is more susceptible to mass wasting and cliff collapse leading to exposures of volatile-rich materials. Finally, a couple of outburst locations appear to coincide with the boundary between Anhur and the northern hemisphere Aker region (Fig. 7d), which also displays a remarkable morphological dichotomy and sharp scarps with associated talus deposits. \par

Interestingly, we did not observe outbursts from any of the large fracture systems (e.g. Wosret's fractured terrains), and only once from a smooth terrain (though notably one of the strongest events, see \#16 in Table 1 and Figure 3). All outburst-related structural features appear in pre-perihelion images, when present (the Wosret pits or the Sobek scarps), and have not been created by the outbursts themselves, but perhaps modified. Over June-September 2016, Rosetta will fly at distances < 10km from the nucleus, which will allow targeted observations of these areas at very high resolution, and a better understanding of their morphologies.\\

An important parameter to retrieve when trying to link dust plumes and local morphology is the angle between the plume and the local surface. This can unfortunately not be achieved with the current data set. As explained in section \ref{sec:source_loc_inv}, we almost never observed the outburst plumes in more than one image. This means that, although we know quite precisely where the source is, we do not have sufficient information to reconstruct the plume in 3 dimensions. This is different than for "usual" jets that can be tracked for many hours and fully inverted. Our best current assumption, from visual clues only, is that the flow is perpendicular to the average local surface. However, this assumption may not hold true for the first few meters, especially if the plume arises from a collapsing cliff. Detailed modeling of flows interactions in the vicinity of the sources may help us constraining the angle of release better in future works.

\begin{figure*}
   \centering
   \includegraphics[width=\hsize]{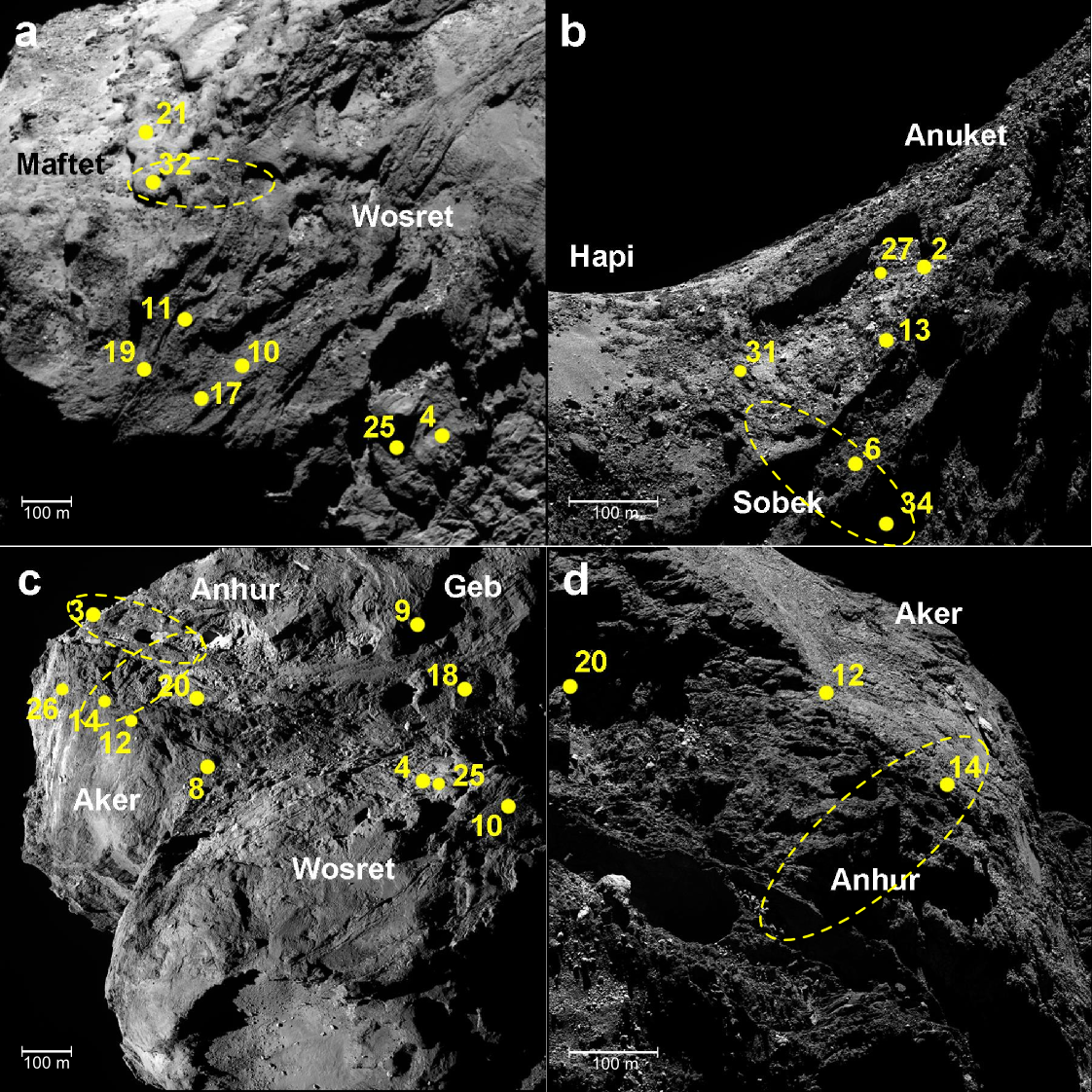}
   \caption{NAC images highlighting morphological and structural settings of a number of candidate outburst locations (numbering consistent with that in Fig. 6 and Table 1) associated with pits and niches in Maftet and Wosret (a), terraced landforms at the boundary between Anuket and Sobek that show extensive debris deposits in the flatter sections (b), the weakly-consolidated Anhur region on the large lobe, which shows many niches, alcoves, boulders and talus deposits (c), and the boundary between Aker and Anhur (d). Images IDs: [a] NAC\_2016-01-27T07.44.41.724Z\_ID10\_1397549800\_F22, [b] NAC\_2016-05-01T21.52.50.787Z\_IDB0\_1397549800\_F22, [c] NAC\_2016-01-23T17.03.47.168Z\_IDB0\_1397549001\_F22, [d] NAC\_2016-05-02T07.16.00.860Z\_IDB0\_1397549900\_F22. Images $a$ and $c$ were acquired from a distance of 76 km (resolution: 1.4~m/px). Images $b$ and $d$ were acquired from a distance of 18 km (resolution: 34~cm/px). Dashed lines indicate the error ellipses for a few sources that could not be well constrained.}
    \label{fig:morpho}
\end{figure*}

%---------------------------%
% --- Diurnal timeline
%---------------------------%
\subsection{Local time of outbursts}\label{sec:local_time}
Knowing the time and location of each outburst, we can calculate the illumination conditions and local time on the comet, in order to understand whether events are more likely to occur under specific temperature conditions. Spacecraft and comet attitudes and trajectories were retrieved with the SPICE library \citep{acton1996}. The local illumination is calculated for the best available shape model of comet 67P: a combination of the model by \cite{preusker2015} for the northern hemisphere, and \cite{jorda2016} for the southern hemisphere. 

\begin{figure}
   \centering
   \includegraphics[width=\hsize]{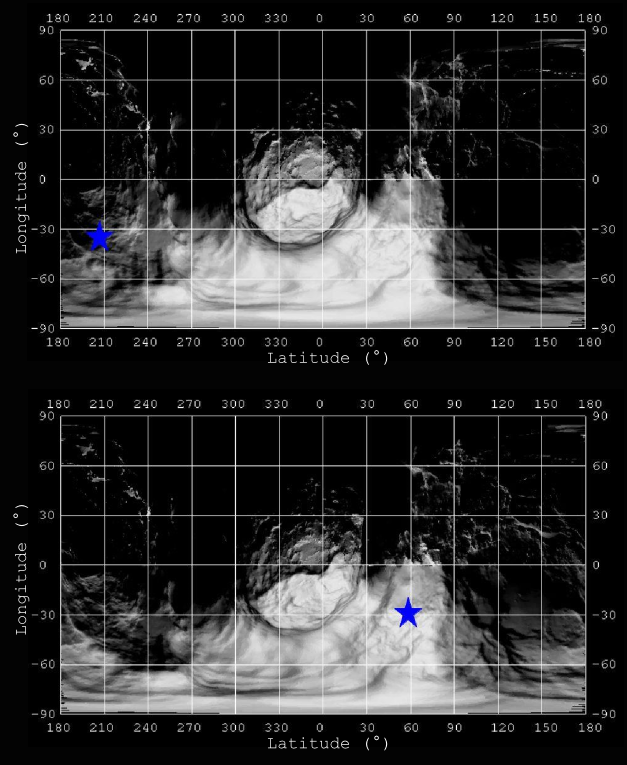}\\
   \caption{Local illumination conditions for outbursts \#7 (top) and \#14 (bottom), examples of an early morning and early afternoon event.}
    \label{fig:local_time}
\end{figure}

Table \ref{tab:outbursts} gives an approximation of the local time since sunrise for each event. This time is calculated from the latitude and longitude offset between outburst source and sub solar location, thus not allowing for variations in shadowing due to the topography. We found that 45\% of the outbursts occurred after about 3 hours of illumination, i.e. shortly after the local noon. The other 55\% of events appear to arise from surfaces that saw their last morning more than 10 hours before. If the comet would be a sphere this would mean night side outbursts. However, due the very complex shape of 67P, one has to look at these events case by case, and indeed they all seem to arise from a very early local morning, as the Sun just starts to shine on the source area. Figure \ref{fig:local_time} shows two examples of local illumination conditions, and illustrates the difficulty of defining the morning terminator.

There is no correlation between type of outburst and local time. The dichotomy is more likely related to a different mechanism (1) Noon outbursts may be linked to buried pockets of volatiles, which need time to get heated enough to trigger an outburst. Shortly after noon is when the local surface reaches its maximum temperature. (2) Early morning outbursts, however, occur almost immediately as the Sun rises. Although the temperature might not yet be too high, the very low thermal inertia ensures that these local times display the steepest temperature gradient. The surface heats up in a few minutes, with a $\Delta T/\Delta t ~(K.s^{-1})$ large enough to trigger thermal cracking (see for instance \cite{alilagoa2015}), and can potentially lead to parts of the surface breaking up.

\cite{belton2008} have estimated the local time of 14 mini-outbursts detected on comet 9P/Tempel 1. Like for 67P, they found a non-random distribution. However, they surprisingly never detected dawn or early morning outbursts, which account for more than half of what Rosetta observed on 67P. Our current interpretation that early morning outbursts are a consequence of rapid changes of temperature leading to cracking of the surface. It may not be happening on 9P because the slow rotation (40 hours) does not allow high thermal stresses to build up. It may also be that 9P's surface has been more processed than 67P's due to its longer time in the inner Solar System, and the current physical/compositional properties of the upper crust differ on both comets.

%---------------------------%
% Discussion
%---------------------------%
\section{Discussion: outbursts mechanisms}\label{sec:mechanism}
The name outbursts itself indicates a violent event, akin to an explosion in the subsurface releasing a large amount of material in a very short time relatively to other forms of activity. One of the most prominent theories invokes the build up of high pressure under the de-volatilized surface layers. The pressure will increase until it overcomes the tensile strength of the surface, at which point the gas pocket will erupt, releasing all the gas and accelerating the surrounding non-volatile material. This will lead to the formation of a small pit or depression \citep{belton2013}, and potentially expose fresh volatile material that will continue to sublime. Such mechanism has been discussed for comet 67P in the case of the Imhotep outburst observed in February 2015 \citep{knollenberg2015}, or the outbursts on 9P/Tempel 1 \citep{belton2008, belton2013}. This process, however, requires that the thermal wave (diurnal or seasonal) is able to reach the depth at which ices are available. Hence, for a region of homogeneous surface properties, most outbursts should occur around the same local time/local solar incidence. As explained in section \ref{sec:local_time}, we do not see this on 67P; the same area can outburst shortly after noon or early in the morning. It implies that this mechanism requires local compositional or physical heterogeneity on a scale comparable to the outburst footprint (<100m), with areas highly enriched in very volatile material such as $CO_2$ ice or amorphous water ice, the latter being often invoked to explain outbursts detected in ground-based observations \citep{prialnik1992}.

On the basis of our observations of the dust plumes and of the local nucleus morphology, we can infer a possible mechanism generating the outbursts.
\cite{vincent2016} has proposed receding fractured cliffs as the major process responsible for the usual dust jets seen around 67P and other comets. In this scheme, small fractures lead to enhanced inwards heat flux and acceleration of the gas in a nozzle-like structure, which form of small jets that can merge into larger feature as they expand away from the surface. \cite{hoefner2015} has shown that fractures are an efficient heat trap, but require specific illumination conditions to achieve full potential, namely the Sun shining directly into the fracture so that the maximum input flux reaches the bottom. This means that local activity can only be sustained by having many small fractures subsequently activated as the nucleus rotates. This is indeed the case in many areas, as discussed in \cite{vincent2016}. But what if the Solar insolation reaches its maximum over a large fracture ? For instance the crack in Hapi region that seems to separate the two lobes of the comet \citep{thomas2015}, or a 500 m long fracture in the Anuket region \citep{elmaarry2015b}, both a few meters wide ? One would expect the same process as for the jets to take place, although enhanced by at least an order of magnitude due to the larger size of those fractures. This high solar input concentrated on a very localized area would lead to an outburst. Unfortunately, although the right illumination conditions are achieved regularly, we have never detected an outburst from a large fracture. This is perhaps not surprising as the heat trap effect would prevent them from retaining much volatile material anyway. The only possibility would be a sudden opening of the fracture either laterally or in depth, which would expose fresh material again.\\

Here we suggest an alternative, new process, which could explain the observed outbursts without the need for local ice reservoirs, or specific illumination conditions. Jet activity from fractured cliffs leads to a weakening of the wall structure until it collapses, a phenomenon observed on 67P and other comets and described in details in \cite{vincent2016}. As most outbursts are located near cliffs presenting evidence of mass wasting, it is tempting to link the two processes. That is to say that with the proposed mechanism, most of the dust is being generated during the collapse, rather than ejected from the surface by an explosion. Of course, we certainly need a gas flow to accelerate the dust away from the surface but the collapse itself may not be triggered by activity. This process requires less energy input than other mechanisms because cometary cliffs are extremely weak (tensile strength < 100 Pa) and any small perturbation can lead to their fall. Additionally, once the dust is released it is easier to accelerate it away because the gas flow does not need to overcome cohesion forces that were keeping the grains on the nucleus surface.

Understanding the details of a cometary cliff collapse is challenging because we do not know well the mechanical properties of the material material. However, it is possible to make a parallel with what happens on Earth. Indeed, as a first approximation the growth and collapse of cliffs is controlled by the ratio between their inner cohesion and the local gravity \citep{melosh2011}. This ratio is the same on Earth and 67P (strength values from \cite{groussin2015a, vincent2015b, bedjaoui2010}):
$$ \textrm{Earth: } \frac{c [Pa]}{g [m.s^{-2}]} = \frac{10^6}{10} = 10^5 \quad [kg.m^{-2}] $$
$$ \textrm{~~67P: } \frac{c [Pa]}{g [m.s^{-2}]} = \frac{50}{5.10^{-4}} = 10^5 \quad [kg.m^{-2}] $$

This means that, as a simplified model, one can look at the behavior of cliff instability on Earth as a proxy for what is taking place on 67P. Upon collapse, terrestrial falling walls like coastal chalk cliffs tend to break not in large chunks, but rather crumble into dust as stresses propagate into an already extremely weak material \citep{mortimore2004}. Note that this is only a qualitative assessment, as Earth cliffs are initially weakened by processes far different than those on the comet, like water erosion at their base or alternating dry weather and heavy rainfall.

The size distribution of material in taluses on 67P shows a consistent lack of large pieces (debris always <10~m, \cite{pajola2015}) which supports a crumbling behavior rather than a break up in many large fragments. As the wall turns into fine pieces, the dust is immediately available to be picked up by the outward flow of gas that surrounds the nucleus. Moreover, if the cliff happens to be illuminated at that time, the sudden exposure of fresh material on its new wall will increase the local gas flux, further enhancing the outward dust transport, and creating a large fan of dust which we can detect as an outburst. 

At least one of our observed outbursts (\#1) is a promising candidate for this mechanism. This event was observed by the NavCam on 10 July 2015, and the plume source appears to be close to the Aswan cliff described in \cite{pajola2016a}. The outburst occurred in the middle of the night, in a region receiving only very few insolation because of polar winter (latitude source: +74\degree, sub-solar latitude: -30\degree).
The source region is particularly interesting because it presents some of the most intriguing topographic features of the comet (active pits, see \cite{vincent2015b}) and strong evidence for regressive erosion by cliff collapse linked to jet activity \citep{vincent2016}. In their morphological analysis of the cliffs, \cite{vincent2016} and \cite{pajola2016a} have presented large fractures on the edge of the cliff, indicating blocks on the verge of falling. These blocks have now fallen, and the cliff presents a different edge since the second half of July 2015, along with a modified talus (full details in Pajola et al, 2016, submitted). Based on our observations of this area (very poorly illuminated at that time), we can date the event to sometime in the first half of July 2015. As we did observe a particularly strong outburst (\#1 in Table \ref{tab:outbursts}) from this specific place in the right epoch, it seems reasonable to consider a relation between the two events.

On 67P, the typical size of the wall chunks threatening to fall, or having already fallen (see examples in \cite{vincent2016, pajola2016a}) is typically a few tens of meters along the edge of the cliff and about 10 meters in the other directions. Let us consider a typical wall segment of $50 \times 10 \times 10~m$. i.e. $5000~m^3$ and $2.35 \times 10^6~kg$. We can estimate how much material ends up on the ground upon collapse by looking at the size distribution of blocks in taluses on the nucleus. \cite{pajola2015} have shown that these blocks are typically smaller than 10~m and their cumulative size distribution follow a power law with a slope between -3.5 and -4. The cumulative size distribution for the Aswan area described above has a power law slope of -3.9, and the largest boulders are 4 blocks of about 5~m diameter. By integrating over the size distribution of the talus, and assuming that it is representative of the latest event in this area, we find that all blocks between 0.5~m (detection limit) and the maximum size of 5~m sum up to a total mass of 4500~$m^3$, i.e. 90\% of the mass of a typical falling cliff fragment. The remaining 10\% are the smaller particles that could be ejected as an outburst plume. Such an event would therefore eject about 500~$m^3$ of cometary material, or 235 tons. This is perfectly in agreement with our mas estimate for a typical outburst: 60-260 tons, see Section \ref{sec:intensity}.

~\\

The original perturbation leading to final collapse of a weakened wall remains an open question. Direct activity from the source is an option, but one may also consider vibrations induced by activity in the vicinity or any other process: a small impact, tidal stress, rotation tress... It is difficult, however, to calculate precisely how efficiently vibrations can propagate in the nucleus, as we do not know its internal structure and mechanical properties. We can only get an estimate of the P and S waves velocities using the textbook relations:
$$v_P = \sqrt{\frac{K+(4/3)\mu}{\rho}} \quad \textrm{and} \quad v_S = \sqrt{\frac{\mu}{\rho}}$$
with $K$ the bulk modulus, $\mu$ the shear modulus, and $\rho$ the material density.

If we take $K \simeq$ compressive strength = $150~Pa$, $\mu \simeq$ tensile strength = $50~Pa$ \citep{groussin2015a}, and $\rho = 470~kg.m^{-3}$ \citep{jorda2016}, we obtain $V_P = 0.68~m.s^{-1}$ and $V_S = 0.33~m.s^{-1}$. 

This velocities are quite low, but comparable to the escape velocity, for instance ($0.8~m.s^{-1}$). We do not know, $Q$, the coefficient of attenuation of seismic waves for cometary material but it is likely to be high considering the fact that waves are traveling in a very porous granular material. Therefore it is reasonable that if cliffs collapse are triggered by vibrations, the source must be located in their close vicinity.

We note that \cite{steckloff2016} have invoked avalanches to explain some of the activity of comet 103P/Hartley 2 and suggested rotational stresses as the trigger. This does not apply to 67P as the outbursts locations do not correlate with the areas of maximum centrifugal force. Additionally, we observe collapse of consolidated material rather than flows of granular material.

%---------------------------%
% CONCLUSION
%---------------------------%
\section{Conclusion}
We have presented a series of 34 transient release of gas and dust by the nucleus of comet 67P over the three months surrounding its perihelion passage mid-August 2015.

We found that outbursts occur about every 2.4 nucleus rotation and last at most a few minutes. 
They are comparable in scale and temporal variation to what has been observed on other comets. 
The dust plumes released by these events can be classified into 3 main morphologies: narrow jets, broad plumes, or a combination of both. 

We produced a map of the source locations of these events, and discussed the associated local topography. We find that the spatial distribution of outbursts locations on the nucleus correlates well with morphological region boundaries, especially areas marked by steep scarps or cliffs. 

Outbursts occur either in the early morning or shortly after the local noon, possibly indicating two types of processes: Morning outbursts may be triggered by thermal stresses linked to the rapid change of temperature, while afternoon events are most likely related to the diurnal or seasonal heat wave reaching volatiles buried deeper in the nucleus than those responsible for the more typical jets. In addition, we propose that some events can be the result of a completely different mechanism, in which most of the dust is released upon the collapse of a cliff.

This idea will be tested with a more detailed morphological study using forthcoming high resolution images, joint analysis with other instruments on board Rosetta, and numerical modeling.

%---------------------------%
% ACKNOWLEDGEMENTS
%---------------------------%
\section*{acknowledgements}
OSIRIS was built by a consortium led by the Max Planck Institut f\"ur Sonnensystemforschung, G\"ottingen, Germany, in collaboration with CISAS, University of Padova, Italy, the Laboratoire d'Astrophysique de Marseille, France, the Instituto de Astrofisica de Andalucia, CSIC, Granada, Spain, the Scientific Support Office of the European Space Agency, Noordwijk, The Netherlands, the Instituto Nacional de Tecnica Aeroespacial, Madrid, Spain, the Universidad Politecnica de Madrid, Spain, the Department of Physics and Astronomy of Uppsala University, Sweden, and the Institut f\"ur Datentechnik und Kommunikationsnetze der Technischen Universit\"at Braunschweig, Germany.

The support of the national funding agencies of Germany (DLR), France (CNES), Italy (ASI), Spain (MINECO), Sweden (SNSB), and the ESA Technical Directorate is gratefully acknowledged.

We thank the Rosetta Science Ground Segment at ESAC, the Rosetta Mission Operations Centre at ESOC and the Rosetta Project at ESTEC for their outstanding work enabling the science return of the Rosetta Mission.

All Rosetta NavCam images are released under the Creative Commons license ESA/Rosetta/NAVCAM - CC BY-SA IGO 3.0

This research has made use of NASA's Astrophysics Data System Bibliographic Services.

The authors wish to thanks M. J. S. Belton for his helpful criticism of the paper during the review process.

%---------------------------%
% BIBLIOGRAPHY
%---------------------------%
\bibliographystyle{mnras}
\bibliography{references}

\bsp	% typesetting comment
\label{lastpage}
\end{document}